\documentclass[vecphys]{svmult}

\usepackage{makeidx}         
\usepackage{graphicx}        
\usepackage{multicol}        
\usepackage[bottom]{footmisc}

\begin{document}
%
\title*{States of Traffic Flow in Deep Lefortovo Tunnel (Moscow): Empirical Data}
%
\author{Ihor Lubashevsky\inst{1}
   \and Cyril Garnisov \inst{2}
   \and Reinhard Mahnke\inst{3}
   \and Boris Lifshits \inst{2}
   \and Mikhail Pechersky\inst{2}
}
\authorrunning{Ihor Lubashevsky et al.}
%
%
\institute{A.M. Prokhorov General Physics Institute of Russian Academy of
Sciences, Vavilov str., 38, Moscow, 119311 Russia \texttt{ialub@fpl.gpi.ru}
    \and
Research and Project Institute for City Public Transport,
Sadovo-Samotechnay,~1, Moscow, 103473 Russia \texttt{mpechersk@tochka.ru}
    \and
Universit\"at Rostock, Institut f\"ur Physik, D--18051 Rostock, Germany
\texttt{reinhard.mahnke@uni-rostock.de}
}
\maketitle              

\section*{Traffic flow in long tunnels}

Traffic flow dynamics in long highway tunnels has been studied individually
since the middle of the last century (see, e.g., Refs~\cite{Tun1,Tun2}).
Interest to this problem is due to several reasons. The first and, may be, main
one is safety. Jam formation in long tunnels is rather dangerous and detecting
the critical states of vehicle flow leading to jam is of the prime importance
for the tunnel operation. However, the tunnel traffic in its own right is also
an attractive object for studying the basic properties of vehicle ensembles on
highways because, on one hand, the individual car motion is more controllable
inside tunnels with respect to velocity limits and lane changing. On the other
hand, long tunnels typically are equipped well for monitoring the car motion
practically continuously along them, which provides a unique opportunity to
receive a detailed information about the spacial-temporal structures of traffic
flow.

\begin{figure}
\begin{center}
\includegraphics[scale = 0.65]{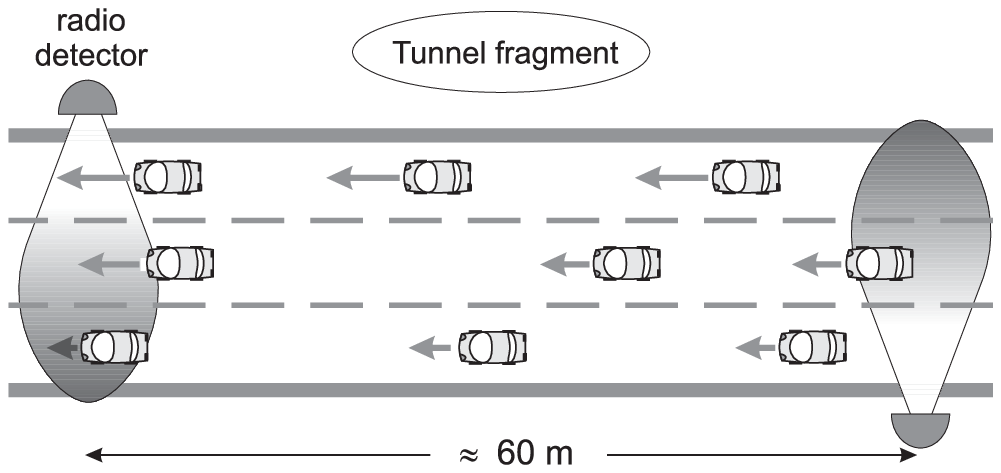}\vspace*{\baselineskip}
\includegraphics[scale = 0.80]{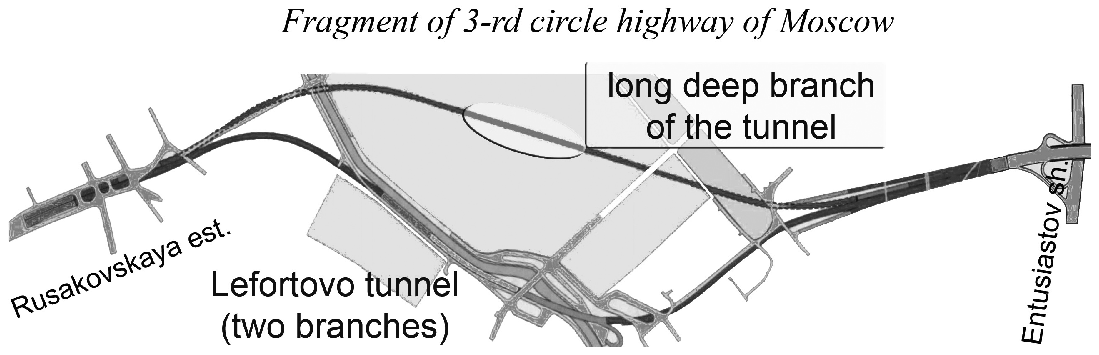}
\end{center}
\caption{Lefortovo tunnel structure.} \label{F:LT}
\end{figure}

By this paper we start analysis of the basic properties exhibited by tunnel
congested traffic that is based on empirical data collected during the last
time in several new deep long tunnels located on the 3-rd circular highway of
Moscow. Here preliminary results for the Lefortovo tunnel (Fig.~\ref{F:LT}) are
presented. It comprises two branches and the upper one is a deep linear three
lane tunnel of length about 3~km. Exactly in this branch the presented data
were collected. The tunnel is equipped with a dense system of stationary
radiodetetors distributed uniformly along it chequerwise at spacing of 60~m.
Because of the detector technical features traffic flow on the left and right
lanes is measured at spacing of 120~m whereas on the middle lane the spacial
resolution gets 60~m. The data were averaged over 30~s.

Each detector measures three characteristics of the vehicle ensembles; the flow
rate $q$, the car velocity $v$, and the occupancy $k$ for three lanes
individually. The occupancy is an analogy to the vehicle density and is defined
as the total relative time during with vehicles were visible in the view region
of a given detector within the averaging interval. The occupancy is measured in
percent.

\section*{Observed cooperative motion of vehicle ensemble}

\begin{figure}
\centering
\includegraphics[width = 100mm]{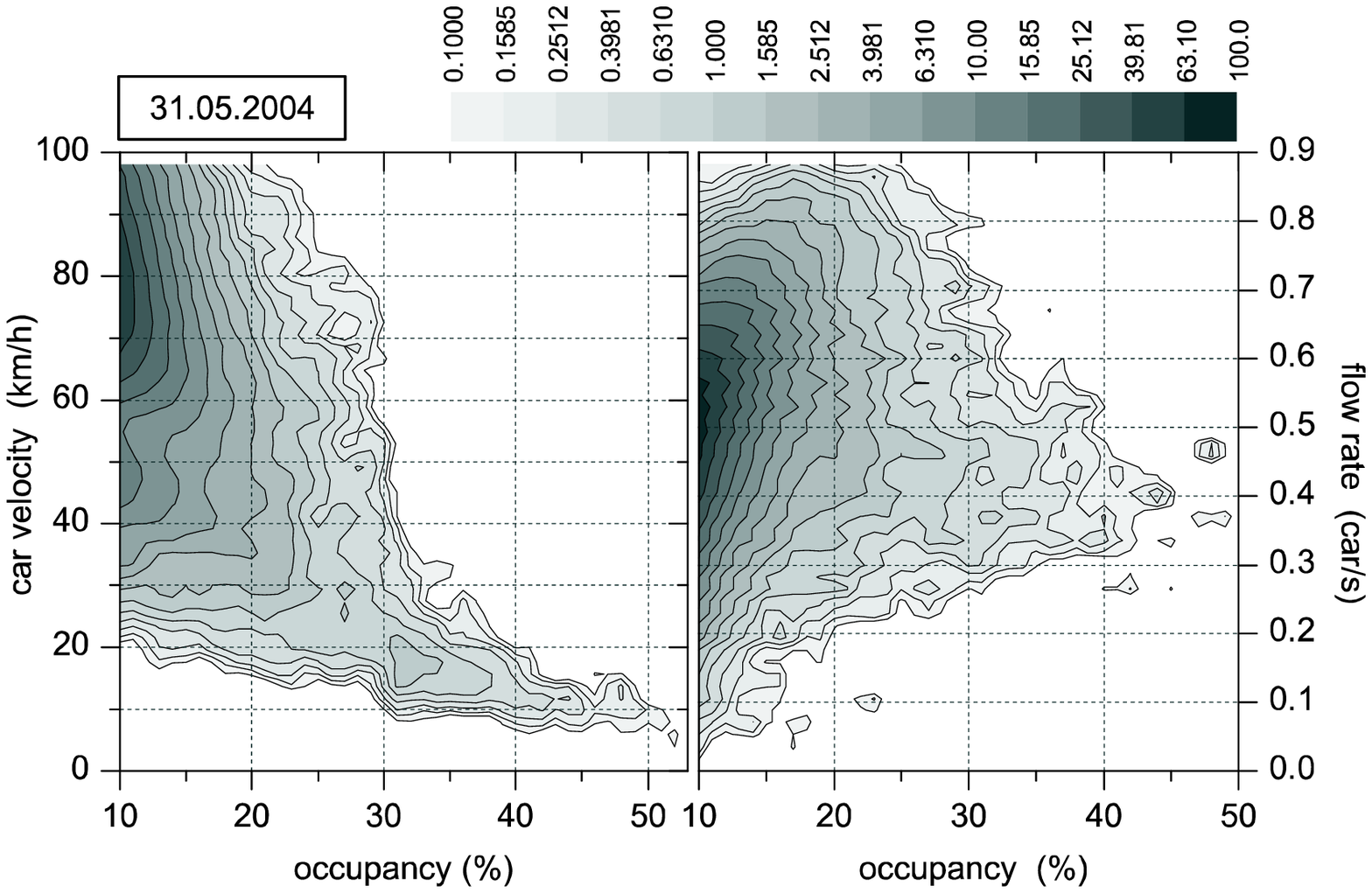}\vspace*{0.5\baselineskip}
\includegraphics[width = 75mm]{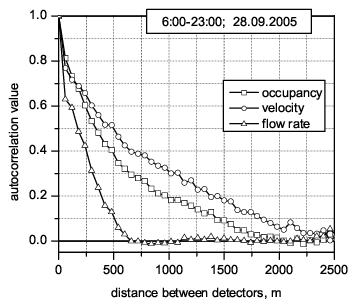}
\caption{Fundamental diagrams (upper frames) and autocorrelations in the
occupancy, car velocity, and flow rate measured by differing detectors vs the
distance between them (lower frame).} \label{F:FD}
\end{figure}

This section demonstrates that the observed traffic flow indeed exhibits
cooperative dynamics when the vehicle density becomes high enough. To do this
figure~\ref{F:FD} (upper frames) depicts the phase planes $\{k,v\}$ and
$\{k,q\}$ with the distribution of the traffic flow states fixed by all the
detectors on 31.05.2004. These phase planes were divided into cells of size
about $1\%\times 2$~km/h and $1\%\times 0.02$~car/s, respectively, and the
number of states measured with frequency 1/30~s$^{-1}$ and falling in a chosen
cell were countered, giving the corresponding distributions. These
distributions in some relative units are represented here in the form of the
level contours. The left side of each window matches the free flow states as
clear seen in the right window, where the darkened region visualizes an upper
fragment of the flow-density relation of the free car motion. However the
obtained distributions even for the free flow are widely scatted which seems to
be due to the essential heterogeneity of the free flow with respect to the
headway distances. The middle parts of these windows visualize another mode of
traffic flow corresponding to the so-called widely scattered states or the
synchronized vehicle motion (for a review see Refs~\cite{H,K}). In fact here
the distribution levels cover rather wide regions and do not follow each other
so frequently as in the left part. Exactly this mode is usually related to the
cooperative vehicle motion.

Figure~\ref{F:FD} (lower frame) exhibits the spatial autocorrelations in the
occupancy, car velocity, and flow rate measured by different detectors at the
middle lane on 28.09.2005 when congested traffic was dominant. In agreement
with the single-vehicle data \cite{SVD} the congested vehicle motion is
characterized by essential correlations especially in the car velocity. The
flow rate measurements are correlated substantially only within several
neighboring detectors (on scales about several hundred meters) whereas the
velocity measurements as well as the occupancy ones are correlated at half of
the tunnel length, i.e. at scales about one kilometer.

\section*{Pattern of vehicle ensemble dynamics in the phase space}

\begin{figure}
\centering
\includegraphics[width = 90mm]{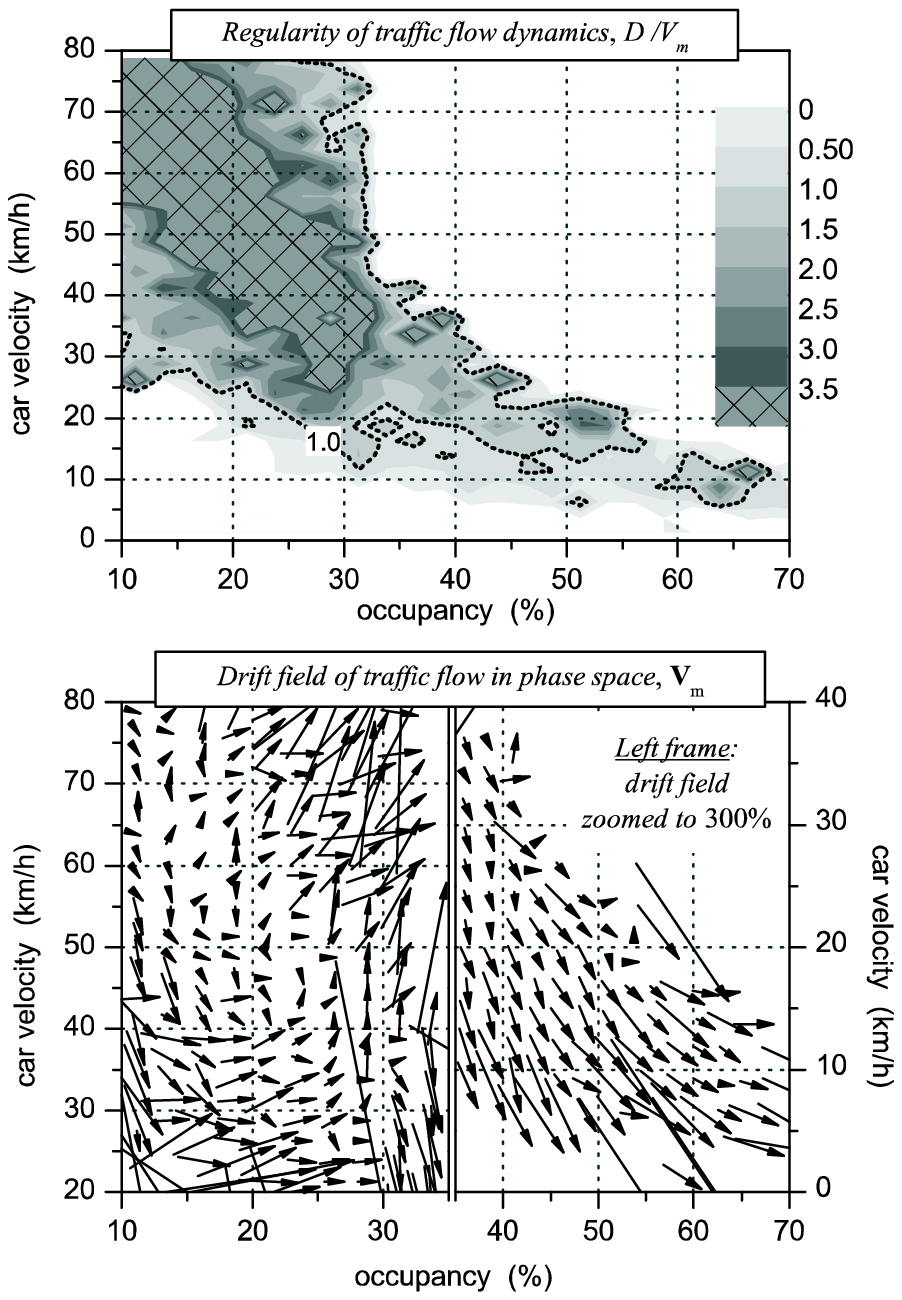}
\caption{Structure of ensemble vehicle dynamics in the phase space $\{o,v\}$.
The upper window visualizes distribution of ratio between the random and
regular components of the effective forces. The lower window depicts the
regular drift field.} \label{F:Pat}
\end{figure}

The characteristics of the vehicle ensemble dynamics in the phase space
$\{k,v\}$ were studied in the following way, replicating actually the technique
of Ref.~\cite{W} used in a similar analysis. The plane $\{k,v\}$ is divided
into cells $\{\mathcal{C}\}$ of size $2.5\%\times2.5$~km/h. Let at time $t$ the
traffic flow measurements of a given detector fall in a cell $\mathcal{C}_i$
and in the averaging time $dt = 30$~s the next measurements of the same
detector are located in a cell $\mathcal{C}_j$. Then the vector
$d\mathbf{r}:=\{dk_t,dv_t\}$ such that $dk_t = k_j -k_i$ and $dv_t = v_j - v_i$
describes the system motion on the phase plane at the given point $\mathbf{r}_i
:= \{k_i,v_i\}$ at time $t$. These vectors were calculated using the data
collected on 28.09.2005 by all the detectors. Averaging the found vectors gives
the drift field $\mathbf{V}_m(\mathbf{r}) = \left<d\mathbf{r}\right>/dt$ and
the intensity $D(\mathbf{r})$ of an effective random force determined as
$$
D dt = \sqrt{\left<\left|d\mathbf{r}\right|\right>^2 -
\left<d\mathbf{r}\right>^2}\,.
$$

Figure~\ref{F:Pat} exhibits these fields. The upper window depicts the ratio
$\eta :=D/|\mathbf{V}|_m$, namely, its variations from 0 up to 3.5. The white
region comprises the cells where no measurements were obtained. The hatched
domain matches the ratio $\eta > 3.5$, where the vehicle ensemble dynamics can
be regarded as pure random. The region between them contains several levels of
the ratio $\eta$ variations and the level $\eta = 1.0$ is singled out in
Fig.~\ref{F:Pat}. For smaller values of $\eta$ the dynamics of vehicle ensemble
becomes practically regular.

The lower window of Fig.~\ref{F:Pat} shows the drift field
$\mathbf{V}_m(\mathbf{r})$. Since its intensity changes essentially at
different parts of the plane $\{k,v\}$ two frames are used to visualize it. In
the left frame the drift field is zoomed in by three times relative to the
right one. Let us consider them individually. The system dynamics in the right
frame is rather regular and the filed $\mathbf{V}_m(\mathbf{r})$ corresponds to
the irrelievable drift of vehicle ensemble to smaller velocities and higher
densities. In other words, it is some visualization of the jam formation. In
fact one or two jams were the case on that day. It should be noted that the
transition region separating the left frame pattern being rather chaotic and
the given one is relatively thin, it is located at $k = 35\%$ and has a
thickness less then 5\%\,. So the observed jam formation seems to proceed via
some breakdown in the cooperative vehicle motion, which is an agreement with
other data \cite{K}.

The pattern shown in the left frame matches the upper one in structure. Inside
a neighborhood $\mathcal{Q}_0$ of the decreasing frame diagonal the traffic
dynamics is practically pure chaotic, at least, the found values of
$\mathbf{V}_m(\mathbf{r})$ are relatively small and their directions do not
form any regular pattern. As it must, outside this domain the field
$\mathbf{V}_m(\mathbf{r})$ becomes more regular and the obtained data enable us
to estimate its characteristic direction. Unexpectedly, it turns out that the
field $\mathbf{V}_m(\mathbf{r})$ crossing this neighborhood does not change its
direction for backward one as it should be if the domain $\mathcal{Q}_0$ has
contained a zero set of the regular field $\mathbf{V}_m(\mathbf{r})$. Such
behavior of a dynamical system can be explained using the notion of dynamical
traps predicting also the existence of a long-lived state multitude as a
consequence of some nonequilibrium phase transitions caused by the human
bounded rationality \cite{BR1,BR2,BR3}.

\section*{Conclusion}

The paper presents a preliminary analysis of traffic flow data collected in the
Lefortovo tunnel located on the 3-rd circular highway of Moscow or, more
rigorously, in its upper linear branch being a deep three lane tunnel of length
about 3 km. The radiodetectors of vehicle motion are distributed chequerwise
along it practically uniformly at spacing of 60~m. The measured data are
averaged over 30~s.

It is shown that the observed tunnel congested traffic in fact exhibits
cooperative phenomena in vehicle motion, namely, there is a region of widely
scatted states on the fundamental diagrams which is related typically to the
appearance of synchronized traffic. Besides, the spatial autocorrelations in
the occupancy, vehicle velocity, and flow rate measured by different detectors
are found to be essential. Especially it concerns the correlations in the
velocity and occupancy, their correlation length gets values about 1~km. The
occupancy data are correlated on substantially shorter scales about 200--300~m.

The phase portrait of the vehicle ensemble dynamics on the occupancy-velocity
plane is also studied. It is demonstrated that there are two substantially
different region on it. One matches actually the cooperative vehicle motion and
contains some kernel where the dynamics is pure chaotic. It is essential that
the found regular drift outside this region does not change the direction when
crossing it. The latter feature is some prompt to applying the concept of
dynamical traps to describing phase transition in congested traffic. The other
part of the phase plane corresponds to the irreversible stage of jam formation.
The two regions are separated by a rather narrow transition layer located at $k
= 35\%$, which demonstrates that the observed jams originated inside a
congested traffic via some breakdown.

\subparagraph{Acknowledgements:} This paper was supported in part by DFG
Project 436 RUS~17/122/04, RFBR Grant 05-01-00723, and Moscow Grant 1.1.258.

\end{document}